\documentclass[10pt,twocolumn,twoside]{IEEEtran}

\IEEEoverridecommandlockouts                             
\usepackage[pdftex]{graphicx}
\graphicspath{{figure/}}
\usepackage{amsmath}
\usepackage{amssymb}
\usepackage{url}

\usepackage{amsthm} 
\usepackage{cite}
\usepackage{color}
\usepackage{epstopdf}

\usepackage{bm}
\usepackage{comment}

\usepackage{algorithm}
\usepackage{algorithmic}

\usepackage{bbm}
\usepackage{enumitem}
\usepackage{mathtools}

\newcommand{\bx}{\bm{x}}

\newcommand{\bP}{\bm{P}}
\newcommand{\bA}{\bm{A}}

\newcommand{\bp}{\bm{p}}

\newcommand{\bc}{\bm{c}}

\newcommand{\bw}{\bm{w}}

\newcommand{\sN}{\mathcal{N}}

\newcommand{\sE}{\mathcal{E}}

\newcommand{\sG}{\mathcal{G}}

\newcommand{\sL}{\mathcal{L}}

\newcommand{\sT}{\mathcal{T}}

\DeclareMathOperator*{\argmin}{argmin}

\allowdisplaybreaks

\allowdisplaybreaks

\title{Enhance Low-Carbon Power System Operation via 
Carbon-Aware Demand Response}

\author{Xin Chen 
	\thanks{X. Chen is with the Department of Electrical and Computer Engineering, Texas A\&M University, USA; email: {xin\_chen@tamu.edu}.}
	\thanks{ 
		The work was supported in part by the Blockchain \& Energy Research Consortium at Texas A\&M University, and in part by the Smart Grid Center, Texas A\&M Engineering
Experiment Station.
} 
}

\begin{document}

\maketitle

\begin{abstract}

As the electrification process advances, enormous power flexibility is becoming available on the demand side, which can be harnessed to facilitate power system decarbonization. Hence, this paper studies the carbon-aware demand response (C-DR) paradigm, where individual users aim to minimize their carbon footprints through the optimal scheduling of flexible load devices. The specific operational dynamics and constraints of deferrable loads and thermostatically controlled loads are considered, and the carbon emission flow method is employed to compute users' carbon footprints with nodal carbon intensities. Then, an optimal power dispatch model that integrates the C-DR mechanism is proposed for low-carbon power system operation, based on the carbon-aware optimal power flow (C-OPF) method. Two solution algorithms, including a centralized Karush–Kuhn–Tucker (KKT) reformulation algorithm and an iterative solution algorithm, 
are developed to solve the bi-level power dispatch optimization model. Numerical simulations on the IEEE New England 39-bus system demonstrate the effectiveness of the proposed methods.

\end{abstract}
 

\section*{Nomenclature}

\addcontentsline{toc}{section}{Nomenclature}

\subsection{Sets and Parameters}
\begin{IEEEdescription}[\IEEEusemathlabelsep\IEEEsetlabelwidth{$V_1,V_2,V_3$}]
\item [$\sN$] Set of nodes.
\item [$\sE$] Set of branches.
\item [$\sN_i\, (\sN_i^+)$] Set of neighbor nodes of node $i$ (that send power to node $i$).
\item [$\sG\, (\sG_i)$] Set of generators (that are located at node $i$).
\item [$\sL\, (\sL_i)$] Set of loads (that are located at node $i$).
\item [$\sT$] Set of time steps with the time interval $\Delta t$.
\item [$w_{g}^G$] Generation  
emission factor of generator $g$. 
\item [$g_{ij}, b_{ij}$]  Conductance, susceptance of branch $ij$. 
\item [$\underline{P}_{g}^{G}, \overline{P}_{g}^{G}$]  Lower, upper limits of active power output of generator $g$. 
\item [$\underline{Q}_{g}^{G}, \overline{Q}_{g}^{G}$]  Lower, upper limits of reactive power output of generator $g$. 

\item [$\underline{R}_{g}^{G}, \overline{R}_{g}^{G}$]  Lower, upper limits of power ramping rate of generator $g$. 
\item [$\underline{V}_{i}, \overline{V}_{i}$]  Lower, upper bounds of voltage magnitude at node $i$. 
\item [$\underline{\theta}_{i}, \overline{\theta}_{i}$]  Lower, upper bounds of phase angle at node $i$.
\item [$\overline{S}_{ij}$] Apparent power capacity of branch $ij$.

\end{IEEEdescription}	

\subsection{Variables}
\begin{IEEEdescription}[\IEEEusemathlabelsep\IEEEsetlabelwidth{$V_1,V_2,V_3$}]
\item [$P_{ij}^i (P_{ij}^j)$] Active power flow of branch $ij$ from node $i$ to node $j$ measured at node $i$ (node $j$). 
\item [$Q_{ij}^i (Q_{ij}^j)$] Reactive power flow of branch $ij$ from node $i$ to node $j$ measured at node $i$ (node $j$). 
\item [$P_{g}^G (Q_{g}^G)$] Active (reactive) power output of generator $g$.
\item [$P_{l}^L (Q_{l}^L)$] Active (reactive) power of load $l$.
 \item [$V_i,\, \theta_i$] Voltage magnitude, phase angle at node $i$.
\item [$w_i$] Nodal carbon intensity of node $i$.
\end{IEEEdescription}

\vspace{2pt}
\noindent
\textbf{Notes}: Notations with a subscript $t$ denote the values at time $t$, and notations with a superscript $*$ denote the corresponding optimal values. For example, $P_{l,t}^{L*}$ denotes the optimal active load value at time $t$ of load $l$.




\section{Introduction}

Human-induced climate change, primarily caused by the escalating concentrations of carbon dioxide and other greenhouse gases (GHGs) in the atmosphere, is posing grave threats to the well-being of billions of people worldwide, while precipitating widespread and irreversible disruption in the natural environment \cite{portner2022climate}.
In 2022, the global GHG emissions reached approximately 57.8 billion tonnes, up to 36\% of which originated from energy systems, and notably, the combustion of fossil fuels for electricity generation accounted for a significant amount of over 13.8 billion tonnes of GHG emissions \cite{carbonclock}. 
As such,  {deep and rapid decarbonization of electric power systems} has been identified as a top priority by IPCC \cite{portner2022climate,2023ar6} to combat climate change and secure a sustainable future for both human society and the natural environment.


As a complement to the decarbonization endeavors on the power generation side, this paper focuses on harnessing the enormous power flexibility available on the \emph{demand} side to decarbonize power systems at scale. 
This is motivated by three main reasons: 1) 
while carbon emissions originate physically from the generation side, it is the electricity consumption that creates the need for power generation and results in emissions;  2) the widespread  {electrification} process
is leading to a significant increase in electricity demand and rapid proliferation of distributed energy resources (DERs) \cite{akorede2010distributed,pudjianto2007virtual}, such as smart buildings, energy storage, electric vehicles, and solar panels, which can be coordinated to support low-carbon system operation; and 3)  an ever-expanding spectrum of electricity users, including companies, municipalities, and organizations,  
have been committed to the pursuit of carbon-free electricity use, such as Google \cite{google247program},  Microsoft \cite{microsoft}, and the global renewable energy corporate initiative ``RE100" \cite{re100}. 
These users are  proactively 
seeking strategies to reduce their own carbon footprints. Moreover, the substantial flexible energy resources from numerous end-users can be leveraged to accelerate the decarbonization process of power systems.

To engage a large number of diverse end-users in grid decarbonization, 
it is essential to measure and quantify the carbon footprints associated with their electricity consumption, known as demand-side carbon accounting \cite{totalCA2021,world2014scope2}.
The prevailing industrial schemes for calculating carbon footprints primarily rely on grid average emission factors over long time horizons and large geographical areas \cite{world2014scope2} or based on clean power market instruments  \cite{brander2018creative}.  As a result, these schemes neglect the physical power grids and power flow delivery and suffer from low geographic and temporal granularity \cite{ourvisionpaper}. To this end, 
the \emph{carbon (emission) flow} method \cite{kang2015carbon,li2013carbon} is introduced for granular power system carbon accounting, which treats carbon emissions as ``\emph{virtual}" network flows embodied in power flows and transmitted 
from generators to users. The mathematical model of carbon flow in electric power networks is built in \cite{kang2012carbon}. Reference \cite{chen2023carbon} 
presents the carbon flow model for lossy networks and rigorously establishes the conditions that ensure the invertibility of the carbon flow matrix. 
The nodal carbon intensities yielded by the carbon flow method can be used to compute the carbon footprints of end-users with high spatial and temporal granularity; see \cite{kang2012carbon} and Section \ref{sec:cef} for details.

A prominent way to harness the power flexibility of numerous end-users \cite{chen2019aggregate} is through demand response (DR) programs \cite{jordehi2019optimisation}. However, most existing DR schemes, including price-based mechanisms \cite{chen2012real,muratori2015residential}  and incentive-based mechanisms \cite{hu2016framework,chen2021online,chen2020online}, focus on enhancing economic efficiency and system reliability, while the essential goals of grid decarbonization and carbon footprint reduction for users are  {inadequately} considered. 
In \cite{shao2019low}, a low-carbon economic dispatch model is developed with the incorporation of the consumption-side emission penalty scheme,  where the power flow tracing technique is employed to compute consumers' carbon emission rates, and a two-level optimization model is formulated to determine consumers’  penalty rates. Reference \cite{zhang2024networked} proposes a multi-agent reinforcement learning framework  for  low-carbon demand management in
distribution networks considering the carbon emission allowance
on the demand side.  In \cite{yang2022price}, the mechanism and model of price-based low-carbon demand response are proposed based on the electricity price, carbon price, and carbon emission intensity. Reference \cite{xu2022event} builds a low-carbon economic dispatch model for multi-regional power systems, considering the reduction of carbon emission obligation on the load side. 
Nevertheless, these schemes generally do not take into account the specific operational dynamics and constraints of heterogeneous load devices.

This paper studies the carbon-aware demand response (C-DR) paradigm where individual end-users seek to reduce their carbon footprints by optimally scheduling flexible load devices. 
The time-varying nodal carbon intensities calculated by the carbon emission flow method are useful signals from the power grid for quantifying users' carbon footprints and guiding low-carbon electricity consumption behaviors. 
Hence, this paper employs the nodal carbon intensities as control signals to incentivize decarbonization actions from end-users and incorporate this C-DR mechanism into the power dispatch process to enhance low-carbon power system operation. The main contributions of this paper are summarized as follows. 
\begin{itemize}
    \item [1)] This paper proposes a C-DR model for the optimal management of heterogeneous flexible load devices, including deferrable loads and thermostatically controlled loads (TCLs). Then, an optimal power dispatch model that integrates the C-DR mechanism is developed  
for power systems 
based on the carbon-aware optimal power flow (C-OPF) method.
\item [2)] The proposed power dispatch model with the C-DR mechanism is formulated as a bi-level optimization problem, and this paper develops two solution algorithms to solve it: 1) a centralized solution algorithm based on the Karush–Kuhn–Tucker (KKT) reformulation, and 2) an iterative solution algorithm that can preserve private information of users' load devices. 

\item[3)]  We implement and test the proposed C-DR-embedded power dispatch scheme via 
numerical simulations on the IEEE New England
 39-bus system. The simulation results demonstrate the performance of the proposed methods and show that the C-DR mechanism can effectively reduce the grid's carbon emissions and operational costs.
\end{itemize}

The remainder of this paper is organized as follows: Section \ref{sec:cef} presents the preliminaries on the carbon flow method. Section \ref{sec:problem} introduces the carbon-aware demand response and power dispatch models. Section \ref{sec:solution} develops the solution algorithms. 
  Numerical tests are conducted in Section \ref{sec:simulation}, and conclusions are drawn in Section \ref{sec:conclusion}.

\section{Preliminaries on Carbon Emission Flow Method and Carbon Accounting} \label{sec:cef}




Consider a power network described by a connected graph $G(\mathcal{N},\sE)$, where $\sN\!:=\!\{1,\cdots, N\}$ denotes the set of nodes and $\sE\subset \sN\!\times\! \sN$ denotes the set of branches.   The carbon (emission) flow method \cite{kang2015carbon}
 conceptualizes carbon emissions from power generators as  {virtual} attachments to the power flow, which are transported
 through power networks to end-users,  forming the virtual carbon flows. The  {carbon (emission) intensity} $w$ is defined as the ratio that describes the amount of carbon flow associated with one unit of power flow. 
The mathematical model of carbon flow  for a power network is built upon two basic principles \cite{kang2015carbon,chen2023carbon}: 
{\setlist{leftmargin=6mm}
\begin{itemize}
    \item  [1)] \emph{Conservation of nodal carbon mass}: At each node $i$,  the total carbon inflows equal the total carbon  outflow.
    \item [2)]  \emph{Proportional sharing principle}:
    At each node $i$,
     the distribution of total carbon inflows among all outflows is proportional to their power flow values.
\end{itemize}}
These two principles imply that all power outflows from a node $i\in\sN$, including branch outflows and the outflows into nodal loads, share the same carbon intensity. This intensity is defined as the \emph{nodal carbon intensity} $w_i$ and formulated as: 
\begin{align}\label{eq:nci}
  w_i  =     \frac{ \sum_{g\in\sG_i}\! w_{g}^{G} P_{g}^{G} + \sum_{k\in \sN_i^+}\! w_k P_{ki}^i}{ \sum_{g\in\sG_i}\!  P_{g}^{G} + \sum_{k\in \sN_i^+}\! P_{ki}^i },\quad   i\in\sN,
\end{align}
which is the ratio of total carbon inflow to the total power inflow at node $i$.  Given the power flow profile $(P_{ij})_{ij\in\sE}$ and generation power and emission factors $(P_i^G, w_i^G)_{i\in\sN}$, one can compute all the nodal carbon intensities $(w_i)_{i\in\sN}$ by
solving a set of carbon flow equations \eqref{eq:nci}. 
Then, the carbon flow  linked to the power flow $P_{ij}$ of branch $ij\!\in\!\sE$, from node $i$ to node $j$, is calculated as $w_iP_{ij}$, 
and the carbon flow associated with the load  $P_l^L$ at node $i$ is calculated as $w_i P_l^L$ for all $l\in\sL_i$. See \cite{kang2015carbon,chen2023carbon} for more explanations on  the carbon flow model.

According to the GHG Protocol\footnote{The GHG Protocol \cite{world2004greenhouse,world2014scope2} developed by the World Resources Institute (WRI)  provides internationally recognized GHG accounting and reporting standards and guidelines, which are widely used in the industry.}   \cite{world2004greenhouse,world2014scope2}, end-users 
account for  
 (Scope 2)  carbon emissions attributed to their electricity consumption, and the carbon flow method can be employed for granular user-side carbon accounting \cite{ourvisionpaper}. 
Consider a time horizon $\sT\!:=\!\{1,2,\cdots,T\}$ with the time gap $\Delta t$. Let user $l$ be the owner of load $P_l^L$ located at node $i$ with $l\in\sL_i$. Using the  
carbon flow method, 
the carbon footprint of user $l$ over the time horizon $\sT$
is  computed as $\Delta t\sum_{t\in\sT} w_{i,t} P_{l,t}^L$.


\section{Problem Formulation} \label{sec:problem}

In this section, we first introduce the carbon-aware demand response model from the perspective of users. Then we present the optimal carbon-aware power dispatch model from the perspective of the power system operators. 

\subsection{Carbon-Aware Demand Response (C-DR)}

Consider the problem setting when individual users aim to reduce their own carbon footprints through demand response. Specifically,
 user $l$ seeks to optimally schedule the active load power $\bP^L_l\!:=\!(P^L_{l,t})_{t\in\sT}$ over time to minimize both the total carbon footprints associated with electricity consumption and the corresponding electricity bill. Moreover,
we consider two main types of flexible loads: 1) deferrable loads, e.g., EV charging and dishwashers, and 2) thermostatically controlled loads (TCLs), e.g., heating, ventilation, and air conditioning (HVAC) systems.  From the perspective of user $l$ who is located at node $i$, namely $l\!\in\!\sL_i$, 
the optimal carbon-aware demand response (C-DR) models are established as follows.  Depending on the load type, the C-DR model for user $l$ can be either \eqref{eq:load1} or \eqref{eq:load2}. 

\subsubsection{Deferrable Load} With the nodal carbon intensity $\bw_i\!:=\!(w_{i,t})_{t\in\sT}$, the optimal load scheduling decision $\bP^{L*}_l\!:=\!(P^{L*}_{l,t})_{t\in\sT}$ is obtained by solving the C-DR model \eqref{eq:load1}:
\begin{subequations}\label{eq:load1}
    \begin{align}
   \bP^{L*}_l(\bw_i) \!:=\! \argmin_{\bP^L_l} &\, \Delta t  \Big(c_e\!\sum_{t\in\sT}w_{i,t}P_{l,t}^L \! +\! \sum_{t\in\sT}p_{i,t}P_{l,t}^L\Big)  \label{eq:load1:obj}\\
       \text{s.t.}\,& \,   \underline{P}_{l,t}^L\leq  P_{l,t}^L\leq \overline{P}_{l,t}^L,  &&\hspace{-32pt} t\in\sT\label{eq:load1:p} \\
       &  \, \underline{E}_{l}^L\leq \Delta_t  \sum_{t\in\sT} P_{l,t}^L\leq \overline{E}_{l}^L,&&\hspace{-32pt} t\in\sT. \label{eq:load1:e}
    \end{align}
\end{subequations}
Here, $c_e$ is the cost parameter of the carbon emission penalty, and $p_{i,t}$ is the given real-time electricity price. $\underline{P}_{l,t}^L$ and $\overline{P}_{l,t}^L$ are the lower and upper limits of the load power, respectively. $ \underline{E}_{l}^L$ and $ \overline{E}_{l}^L$ are the minimum and maximum total energy required to complete the task, respectively. The objective function in \eqref{eq:load1:obj} is comprised of two parts: the first part is the cost of the total carbon footprints associated with electricity consumption, and the second part denotes the electricity bill cost. These two parts can be balanced by
 adjusting the cost parameters in the objective \eqref{eq:load1:obj}. For example, user $l$ focuses on minimizing the total carbon footprints when letting $p_{i,t}$ be zero, and aims to reduce electricity cost when $c_e$ is set as zero.

\subsubsection{Thermostatically Controlled Load} Similar to model \eqref{eq:load1}, the C-DR optimization model of TCLs is built as \eqref{eq:load2}:
\begin{subequations}\label{eq:load2}
    \begin{align}
        & \text{Equations \eqref{eq:load1:obj}, \eqref{eq:load1:p}}, \\
        & T_{l,t}^{in} = T_{l,t\!-\!1}^{in} + \alpha_l(T_{l,t}^{out}-T_{l,t\!-\!1}^{in}) +\beta_lP_{l,t}^L, &&  t\in\sT \label{eq:load2:dyn}\\
        & \underline{T}_l \leq T_{l,t}^{in}\leq \overline{T}_l, &&  t\in\sT.\label{eq:load2:t}
    \end{align}
\end{subequations}
Here, $T_{l,t}^{in}$ and $T_{l,t}^{out}$ and the inside and outside temperatures at time $t$, respectively. $\alpha_l$ and $\beta_l$ are the parameters that specify the heat transfer characteristics of the environment and the thermal efficiency of the TCL appliance. $\underline{T}_l$ and $\overline{T}_l$ denote the lower and upper bounds of the comfort temperatures. See \cite{li2011optimal} for more explanations.



The nodal carbon intensity $w_{i,t}$ in \eqref{eq:load1:obj} is an effective signal for guiding users' low-carbon electricity consumption, which is
calculated using the carbon flow method outlined in Section \ref{sec:cef}. The nodal carbon intensities are time-varying and depend on
 the generation fuel mix and power flow profiles that change over time. The functional form $ \bP^{L*}_l(\bw_i) $  in \eqref{eq:load1:obj} captures the relation that describes user $l$'s load demand in response to the nodal carbon intensity. In this way, the C-DR models \eqref{eq:load1} and \eqref{eq:load2}   incentivize users to 
 increase (decrease) electricity consumption when the grid exhibits low (high) carbon intensity, fostering 
 spontaneous decarbonization behaviors among numerous users. Since the demand  $\bP^{L*}_l$ relies on the nodal carbon intensity $\bw_i$ that is determined by the grid's power dispatch schemes, we incorporate the C-DR mechanism $\bP^{L*}_l(\bw_i)$ into the power dispatch process in the next subsection.

\subsection{Optimal Carbon-Aware Power Dispatch with C-DR}

Based on the  Carbon-aware Optimal Power Flow (C-OPF) methodology introduced in \cite{chen2023carbon}, we develop the Carbon-aware Power Dispatch (C-PD) optimization model  \eqref{eq:cpd} to optimally schedule various generators, considering C-DR mechanisms. 
\begin{subequations}\label{eq:cpd}
    \begin{align}
         \text{Obj.} \! &\, \min_{P_{g,t}^G,Q_{g,t}^G}\, \sum_{t\in\sT}  \sum_{g\in\sG}\Delta t\Big( f_g(P_{g,t}^G) + c_E\cdot w_{g,t}^GP_{g,t}^G\Big) \label{eq:cpd:obj} \\
           \text{s.t.} &\,   P_{ij,t}^i = \big(  V_{i,t}^2  - V_{i,t}  V_{j,t}\cos(\theta_{i,t} \!-\!\theta_{j,t})\big)g_{ij} \nonumber\\
         &\qquad  - V_{i,t} V_{j,t}\sin(\theta_{i,t}\!-\!\theta_{j,t})b_{ij}, &&\hspace{-43pt}  ij\!\in\!\sE,t\!\in\!\sT   \label{eq:cpd:apf}  \\
        & Q_{ij,t}^i =    \big( V_{i,t}  V_{j,t}\cos(\theta_{i,t} \!-\!\theta_{j,t})- V_{i,t}^2  \big)b_{ij}\nonumber \\
         &\qquad     - V_{i,t}  V_{j,t}\sin(\theta_{i,t} \!-\!\theta_{j,t})g_{ij}, &&\hspace{-43pt}  ij\!\in\!\sE,t\!\in\!\sT   \label{eq:cpd:rpf}  \\
           & \sum_{g\in\sG_i}\! P_{g,t}^G \!-\! \sum_{l\in\sL_i}\! P_{l,t}^{L*}(\bw_i)\! =\!  \sum_{ij\in\sE} \! P_{ij,t}^i, &&\hspace{-43pt}  i\!\in\!\sN,t\!\in\!\sT  \label{eq:cpd:p}\\
        &\sum_{g\in\sG_i}\! Q_{g,t}^G \!-\! \sum_{l\in\sL_i}\!\eta_l P_{l,t}^{L*}(\bw_i)\! = \! \sum_{ij\in\sE} \!Q_{ij,t}^i,&&\hspace{-43pt}  i\!\in\!\sN,t\!\in\!\sT \label{eq:cpd:q} \\
         & \underline{P}_{g,t}^{G} \!\leq\! {P}_{g,t}^{G} \!\leq\! \overline{P}_{g,t}^{G},\, \underline{Q}_{g,t}^{G} \!\leq\! {Q}_{g,t}^{G} \!\leq\! \overline{Q}_{g,t}^{G}, &&\hspace{-43pt}  g\!\in\! \sG, t\!\in\!\sT \label{eq:cpd:gen}
        \\
        & \underline{R}_{g}^{G} \leq {P}_{g,t}^{G} - {P}_{g,t-1}^{G} \leq \overline{R}_{g}^{G}, && \hspace{-43pt}  g\!\in\! \sG, t\!\in\!\sT \label{eq:cpd:ramp}\\
          & \underline{V}_i \leq V_{i,t}\leq \overline{V}_i, \    \underline{\theta}_i\leq   \theta_{i,t}\leq \overline{\theta}_i, &&\hspace{-43pt} i\!\in\! \sN, t\!\in\!\sT   \label{eq:cpd:vol}  \\
          & (P_{ij,t}^i)^2 + (Q_{ij,t}^i)^2 \leq (\overline{S}_{ij})^2, && \hspace{-43pt}  ij\!\in\!\sE,t\!\in\!\sT  \label{eq:cpd:line}\\ 
         &  w_{i,t} (\sum_{g\in\sG_i}\! P_{g,t}^G +  \sum_{j\in \sN_i}  \hat{P}_{ji,t}^i  ) \nonumber \\
         &\qquad = \sum_{g\in\sG_i} \! w_{g,t}^{G} P_{g,t}^{G}\! +\!\! \sum_{j\in \sN_i}\!\! w_{j,t} \hat{P}_{ji,t}^i,&&  \hspace{-43pt} i\!\in\! \sN, t\!\in\!\sT \label{eq:cpd:cf}\\
                    & \hat{P}_{ij,t}^i\! - \!\hat{P}_{ji,t}^i \!=\!  P_{ij,t}^i,   \hat{P}_{ij,t}^i\!\geq\! 0,   \hat{P}_{ij,t}^j\!\geq\! 0,  && \hspace{-43pt}  ij\!\in\!\sE,t\!\in\!\sT \label{eq:cpd:Pij} \\ 
           &  \hat{P}_{ji,t}^i\cdot \hat{P}_{ij,t}^i    =0,\,  && \hspace{-43pt}  ij\!\in\!\sE,t\!\in\!\sT. \label{eq:cpd:com}
    \end{align}
\end{subequations}

The objective \eqref{eq:cpd:obj} aims to
schedule the generation power outputs
to
minimize the total generation cost and generation emission penalty. $f_g(\cdot)$ denotes the generation cost function of generator $g$, such as  the quadratic generation cost function $f_g(P_{g,t}^G)\!:=\! c_{g,2}(P_{g,t}^G)^2 + c_{g,1}P_{g,t}^G +  c_{g,0} $ with the parameters $c_{g,2},c_{g,1},c_{g,0}$. $c_E$ is the carbon emission cost parameter, which accounts for the externalities and regulatory penalties associated with carbon emissions. Equations \eqref{eq:cpd:apf} and \eqref{eq:cpd:rpf} are the power flow equations. Equations \eqref{eq:cpd:p} and \eqref{eq:cpd:q} denote the nodal power balance constraints, where the C-DR mechanism $P_l^{L*}(\bw_i)$ is incorporated, and 
parameter $\eta_l$ indicates a fixed power factor for load adjustment. Equations \eqref{eq:cpd:gen} and \eqref{eq:cpd:ramp} are the generation power capacity and ramping constraints, respectively. Equation \eqref{eq:cpd:vol} denotes the voltage and phase constraints, and equation \eqref{eq:cpd:line} denotes the line thermal constraints. Equation \eqref{eq:cpd:cf} represents the carbon flow equations \eqref{eq:nci}. 
As indicated by the set $\sN_i^+$, the original carbon flow equations $\eqref{eq:nci}$ 
need to pre-determine the power flow directions for all branches to identify the power inflows for each node. However,  the directions of branch power flows are typically unknown prior to solving optimal decision models. To address this issue, we adopt the reformulation technique proposed in \cite{chen2023carbon} and introduce two 
non-negative power flow variables $\hat{P}_{ji,t}^i$ and $\hat{P}_{ij,t}^i$ to replace each branch flow $P_{ij,t}^i$ with  $\hat{P}_{ij,t}^i \!-\! \hat{P}_{ji,t}^i$ in \eqref{eq:cpd:Pij}. 
Specifically, $\hat{P}_{ji,t}^i$ and $\hat{P}_{ij,t}^i$ denote  the power flow components from node $j$ to node $i$ and from node $i$ to node $j$, respectively.
In addition, the complementarity constraint \eqref{eq:cpd:com} is added to ensure that either $\hat{P}_{ji,t}^i$ or $\hat{P}_{ij,t}^i$ must be zero. In this way, the reformulation \eqref{eq:cpd:cf} is equivalent to the original carbon flow equation \eqref{eq:nci}, while the necessity of knowing 
 the branch power flow directions is circumvented; see \cite{chen2023carbon} for more details.

 The above C-PD model \eqref{eq:cpd} 
 integrates the C-DR mechanism $\bP_l^{L*}(\bw_i)$, which
 captures the load demand behaviors of end-users in response to the grid's emission status based on the real-time nodal carbon intensities.
  Moreover, the C-PD model \eqref{eq:cpd} is applicable to both transmission and distribution systems, and other dispatchable energy resources such as energy storage systems can also be incorporated.   
 However, each load   $P_{l,t}^{L*}(\bw_i)$ in equations \eqref{eq:cpd:p} and \eqref{eq:cpd:q} is an  implicit function of the corresponding nodal carbon intensity $\bw_i$  that is 
 defined in the C-DR models \eqref{eq:load1}, \eqref{eq:load2}. As a result, the C-PD  model \eqref{eq:cpd} is formulated as
 a bi-level optimization model \cite{sinha2017review}, which is
 challenging to optimize. Hence, we develop solution algorithms to solve the C-PD model \eqref{eq:cpd} in the next section.

\section{Solution Algorithms for Bi-level C-PD Model} \label{sec:solution}

The C-PD optimization model \eqref{eq:cpd} is formulated as a bi-level optimization problem, 
where the power dispatch optimization contains the load scheduling optimization in the constraints. Bi-level optimization problems \cite{talbi2013taxonomy} are common in practice but are
 generally challenging to solve. A number of approaches have been proposed to solve bi-level optimization, including KKT reformulation \cite{shi2005extended}, descent
methods \cite{savard1994steepest}, penalty function methods \cite{lv2007penalty}, meta-heuristic methods \cite{li2006hierarchical,sinha2014improved}, and others; see \cite{sinha2017review} for a review on bi-level optimization methods.
In this section, we introduce a centralized solution method and an iterative solution algorithm to solve the bi-level C-PD optimization model \eqref{eq:cpd}. 

\subsubsection{Centralized KKT Reformulation Solution Method} 
The C-DR optimization models \eqref{eq:load1} and \eqref{eq:load2} for user $l\in\sN_i$ are linear programming problems  and can
 be equivalently expressed as the uniform matrix form \eqref{eq:load:m}: 
 \begin{subequations} \label{eq:load:m}
     \begin{align}
     \bP_l^{L*}(\bw_i):= &\argmin_{\bP_l^L}\ \bc_i^\top \bP_l^L \qquad\\
        &\, \text{s.t.} \, \bA_l  \bP_l^L\leq \bm{b}_l. \label{eq:load:m:con}
     \end{align}
 \end{subequations}
Here, $\bc_i\! :=\! \Delta tc_e\bw_i +\Delta t \bp_i$ with $\bp_i\!:=\!(p_{i,t})_{t\in\sT}$. Equation \eqref{eq:load:m:con} represents constraints \eqref{eq:load1:p}, \eqref{eq:load1:e} for deferrable loads or constraints \eqref{eq:load1:p}, \eqref{eq:load2:dyn}, \eqref{eq:load2:t} for TCLs with the parameter matrix $\bA_l$ and vector $\bm{b}_l$. 
Assuming that Slater's condition holds for \eqref{eq:load:m}, 
the optimal solutions $\bP_l^{L*}$ of \eqref{eq:load:m} satisfy the KKT conditions \eqref{eq:KKT}: \cite{boyd2004convex}
\begin{subequations}\label{eq:KKT}
    \begin{align}
        & \bc_i + \bA_l^\top \bm{\lambda}_l = \bm{0}, \label{eq:KKT:pri} \\
         & \bm{\lambda}_l^\top (\bA_l  \bP_l^L- \bm{b}_l)  = \bm{0},   \\
        & \bA_l  \bP_l^L\leq \bm{b}_l, \, \bm{\lambda}_l\geq \bm{0}, 
    \end{align}
\end{subequations}
where $\bm{\lambda}_l$ is the dual variable associated with constraint \eqref{eq:load:m:con}.

Hence, 
in terms of the C-PD model \eqref{eq:cpd}, one can substitute
the function $P_{l,t}^{L*}(\bw_i)$ in \eqref{eq:cpd:p} and \eqref{eq:cpd:q} with a variable $P_{l,t}^{L}$ and add the KKT conditions \eqref{eq:KKT} as additional constraints to the C-PD model \eqref{eq:cpd} to make this substitution equivalent. As a result, the reformulated C-PD model \eqref{eq:cpd}, \eqref{eq:KKT} becomes a regular nonconvex optimization problem, which can be solved using the  IPOPT solver \cite{biegler2009large}. Note that in this reformulated C-PD model, $P_{l,t}^{L}, \bm{\lambda}_l$, and $\bw_i$ (embedded in $\bc_i$ in \eqref{eq:KKT:pri})
are all optimization variables. In this way, the power grid dispatch center solves the reformulated C-PD model \eqref{eq:cpd}, \eqref{eq:KKT} to make the optimal generation decisions.

\subsubsection{Iterative Solution Algorithm}

The KKT conditions \eqref{eq:KKT} encompass critical private information regarding the load of each user $l$, and this information is often inaccessible to the power system dispatch center in practice. Therefore, we develop an iterative algorithm, i.e., Algorithm \ref{alg:ite}, to solve the C-PD model. Essentially, Algorithm \ref{alg:ite} alternates between solving the C-PD model (in Step \ref{step:cpd}) and solving the C-DR model (in Step \ref{step:cdr}) with updated load profiles and nodal carbon intensity values.
To enhance convergence and 
address the oscillation issue,  
we introduce the value $f_{cpd}^{opt}$ to store the best C-PD objective value achieved across all iterations, along with $\bP_l^{L,opt}$ to store the corresponding optimal load profile. Furthermore, constraint \eqref{eq:addc} is added to the C-DR model \eqref{eq:load1} and \eqref{eq:load2} to enforce proximity between the load profile decision $\bP_l^{L}$ and the optimal load profile  $\bP_l^{L,opt}$, and the upper bound $M_l/k^{\alpha_l}$ shrinks as the iteration count $k$ increases. For each load $l$, parameters $M_l>0$ and $\alpha_l>0$ are utilized to regulate the shrinkage rate. A key merit of 
 Algorithm \ref{alg:ite} is that users do not need to share the private configuration information of their load devices with the dispatch center but only send their tentative load values $\bP_l^{L,k+1}$ in each iteration $k$. Moreover, in Step 5 of Algorithm \ref{alg:ite}, 
 we can further enhance performance by incorporating the Model Predictive Control (MPC) method \cite{chen2019aggregate}. Specifically, once Algorithm 1 terminates,  only the generation decision for the first time step is executed, followed by advancing the time horizon $\sT$ one step forward. This enables the utilization of updated system information.

 \begin{algorithm} 
 \caption{Iterative Solution Algorithm for C-PD  \eqref{eq:cpd}.}
 \begin{algorithmic}[1] \label{alg:ite}
 \renewcommand{\algorithmicrequire}{\textbf{Input:}}
 \renewcommand{\algorithmicensure}{\textbf{Output:}}

  \STATE \textbf{Initialization.} For each load $l\!\in\!\sL$, initialize load profile $\bP_l^{L,0}\!:=\!(P_{l,t}^{L,0})_{t\in\sT}$, optimal load profile $\bP_l^{L,opt} \!\leftarrow\! \bP_l^{L,0}$, and parameters $M_l\!>\!0$ and $\alpha_l\!>\!0$. Initialize the optimal C-PD objective value $f_{cpd}^{opt}$ with a sufficiently large value. Set the 
  maximum iteration number $K_{\max}$, iteration count $k\leftarrow 0$, and convergence tolerance $\epsilon$.

  \STATE \textbf{Solution of C-PD.} The power system dispatch center: \label{step:cpd}
  {\setlist{leftmargin=4mm}
  \begin{itemize}
      \item  Replace the function $P_{l,t}^{L*}(\bw_i)$  by the load value $P_{l,t}^{L,k}$ in
  \eqref{eq:cpd:p}, \eqref{eq:cpd:q} for all $l\!\in\!\sL, t\!\in\!\sT$;
  \item   Solve the C-PD model \eqref{eq:cpd} to obtain 
  the  generation decision $\bx_{k+1}^{*}:=(P_{g,t}^{G*},Q_{g,t}^{G*})_{g\in\sG,t\in\sT}$, the nodal carbon intensities $\bw_i^*$ for  all $i\in\sN$, and the corresponding optimal objective function value $f_{cpd}^{k+1,*}$.

  \item  Send $\bw_i^*$ to the loads $l\in\sL_i$ at node $i$ for  all $i\in\sN$. 
  \end{itemize}
  }

   \STATE \textbf{Update optimal C-PD objective and load profile.} 
   
  If $f_{cpd}^{k+1,*} < f_{cpd}^{opt} $, let $f_{cpd}^{opt}\leftarrow f_{cpd}^{k+1,*}$ and $\bP_l^{L,opt} \leftarrow \bP_l^{L,k}$. 
  
  \STATE \textbf{Parallel solution of C-DR.} Each load $l\!\in\!\sN_i$: \label{step:cdr}
 {\setlist{leftmargin=4mm}
  \begin{itemize}
      \item  Replace $\bw_i$ with the broadcast nodal carbon intensity value $\bw_i^*$ in \eqref{eq:load1:obj};
   \item  Solve its own C-DR model \eqref{eq:load1} (or model \eqref{eq:load2}) with an additional constraint  \eqref{eq:addc}:
   \begin{align} \label{eq:addc}
       ||\bP_l^{L} - \bP_l^{L,opt}  || \leq M_l/k^{\alpha_l}, 
   \end{align}
   to obtain its optimal load decision $\bP_l^{L*}$;
   \item Set $\bP_l^{L,k+1} \leftarrow \bP_l^{L*}$ and send $\bP_l^{L,k+1}$ to the system dispatch center.
    \end{itemize}
  }
    \STATE \textbf{Termination check.} If $||\bx_{k+1}^{*}-\bx_{k}^{*}||\leq \epsilon$ or $k>K_{\max}$, terminate; and output the generation decision $\bx_{k+1}^{*}$.

    
    
    Otherwise, let $k\leftarrow k+1$ and go back to Step 2. 
  
 \end{algorithmic} 
 \end{algorithm}


\section{Numerical Tests}\label{sec:simulation}

In this section, we first introduce the test system and simulation settings. Then, we present the test results of the C-PD scheme, comparing it with the power dispatch scheme that does not consider the C-DR mechanism.

\subsection{Test System and Settings}

 \begin{figure}
    \centering
    \includegraphics[scale=0.39]{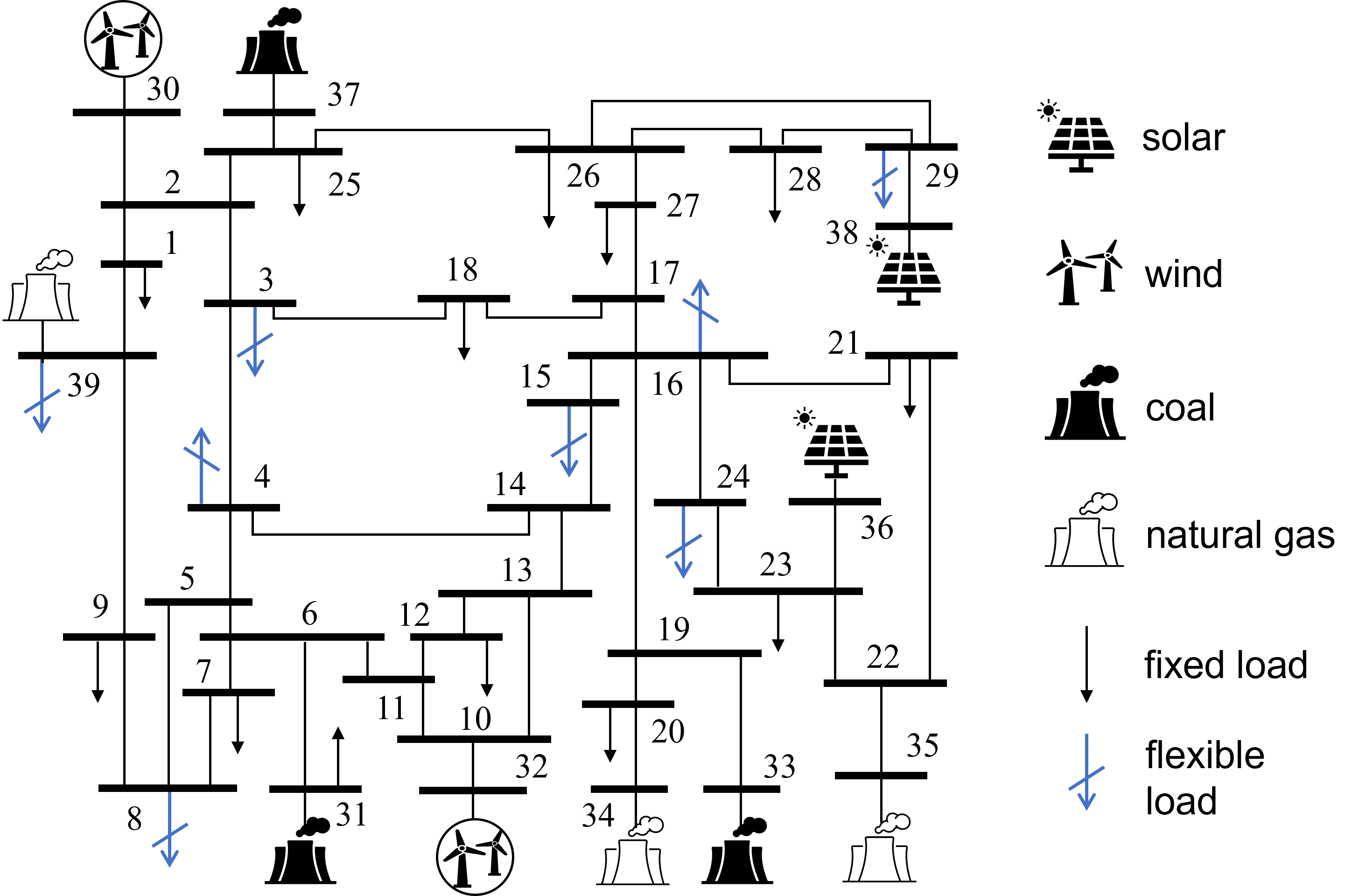}
    \caption{The modified IEEE New England 39-bus test system.}
    \label{fig:39bus}
\end{figure}

\begin{figure}
    \centering
    \includegraphics[scale=0.25]{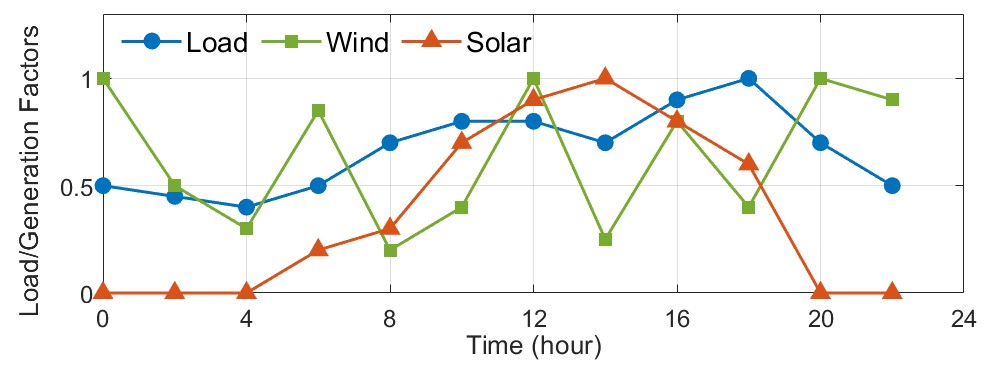}
    \caption{Typical daily curves of solar and wind generation factors and load variation factors for fixed loads.}
    \label{fig:factor}
\end{figure}

Consider day-ahead power dispatch with $T=12$ time steps and 2-hour intervals. 
The modified IEEE New England 39-bus  system, as shown in Figure \ref{fig:39bus}, is used as the test system, including 3 coal power plants, 3 natural gas power plants, 2 wind farms, 2 solar farms, and 21 loads. The loads at buses 3, 4, 8, 15, 16, 24, 29, 39 
are designated as deferrable loads for demand response and are subject to the C-DR model \eqref{eq:load1}. 
The load parameters are set as $\underline{P}_{l,t}^L = 0.5 {P}_{l,t}^{L0}$, $\overline{P}_{l,t}^L = 1.5 {P}_{l,t}^{L0}$, $\underline{E}_{l}^L = \Delta_t  \sum_{t\in\sT} P_{l,t}^{L0}$, and $P_{l,t}^{L0}= \ell_t\cdot P_{l}^{L0}$, where $P_{l}^{L0}$ 
is the nominal load value of load $l$. 
$\ell_t$ is the load variation factor at time $t$, whose values for fixed loads are shown as Figure \ref{fig:factor}, while we let $\ell_t=1$ for flexible loads to facilitate the comparison of simulation results. 
In the numerical tests, users are assumed to focus on minimizing their carbon footprints with cost parameters $c_e = 1$ and $p_{i,t}=0$ in \eqref{eq:load1:obj}, to highlight the decarbonization performance.
The generation carbon emission factors $w_{g,t}^{{G}}$ are 2.26, 0.97, and 0 (lbs/kWh) for coal plants, natural gas plants, and renewable generators, respectively. 
The daily load variation factors $\ell_t$ and the renewable generation factors $\mu_{g,t}$ are shown as Figure \ref{fig:factor}. 
The actual value of each load at each time is determined by multiplying the daily load variation factor with the nominal load value. 
The wind and solar generators are dispatchable, and their maximum power $\overline{P}_{g,t}^{G} = \mu_{g,t}\cdot C_{g}^G $ with the installation capacity $C_{g}^G$. 
The nominal loads, generation capacities, and other system parameters are provided in \cite{ieee39}.

\subsection{Simulation Results}

\begin{figure}
    \centering
    \includegraphics[scale=0.25]{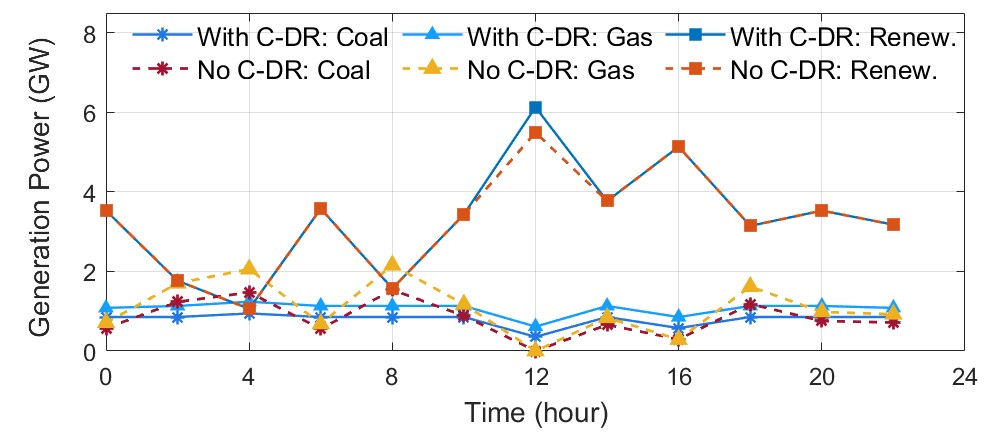}
    \caption{Optimal generation profiles over time under the power dispatch schemes with and without the integration of the C-DR mechanism.}
    \label{fig:gen}
\end{figure}

\begin{figure}
    \centering
    \includegraphics[scale=0.25]{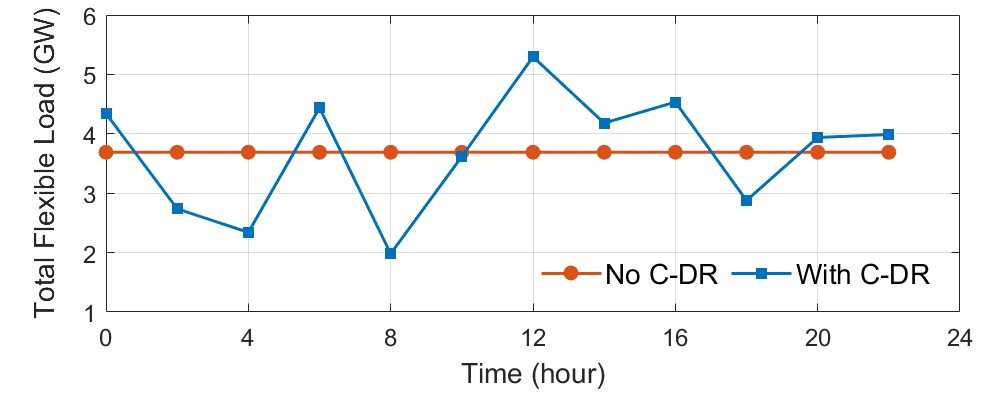}
    \caption{Total flexible load profiles over time under the power dispatch schemes with and without the integration of the C-DR mechanism.}
    \label{fig:load}
\end{figure}

In the simulation tests, we evaluate the effectiveness of the proposed C-PD scheme  \eqref{eq:cpd} with the C-DR mechanism, contrasting it with a power dispatch scheme that does not incorporate the C-DR mechanism. 
The generation decisions $P_{g,t}^{G*}$  of these two schemes for different types of generations are shown in Figure \ref{fig:gen}. It is seen that the natural gas and coal generation profiles under the C-DR-embedded power dispatch scheme exhibit a flatter and smoother trend compared to the scheme without considering C-DR. In addition, at 12 pm, the C-DR-embedded power dispatch scheme achieves higher total renewable generation (i.e., solar and wind), with no curtailment of renewable generation due to the adjustment of flexible loads.  
Figure \ref{fig:load} illustrates the total flexible load decisions $\sum_{l\in\sL_{fle}}\!P_{l,t}^{L*}$ over time.
Since $\underline{E}_{l}^L = \Delta_t  \sum_{t\in\sT} P_{l,t}^{L0}$ is set for all flexible loads, both power dispatch schemes need to supply the same amount of total load energy. From Figures \ref{fig:factor} and \ref{fig:load}, it is observed that under the C-DR mechanism, users spontaneously adapt their flexible loads, aligning them with the temporal variations in carbon emission intensity (or renewable generation). For instance, the peak total flexible load occurs at 12 pm when both solar and wind generation are high; the load nadirs observed at 4 am, 8 am, and 6 pm correspond to the low wind and solar generations during these periods.
This adjustment pattern enables users to minimize their carbon footprints by strategically shifting load consumption to periods characterized by lower carbon intensity. 
As a result, due to the integration of the C-DR mechanism, the total system operational cost is reduced from $13.7\times 10^6$ \$ to  $10.7\times 10^6$ \$, and the overall system emissions are curtailed from $70.08$ Mlbs to $68.34$ Mlbs.

\begin{figure}
    \centering
    \includegraphics[scale=0.25]{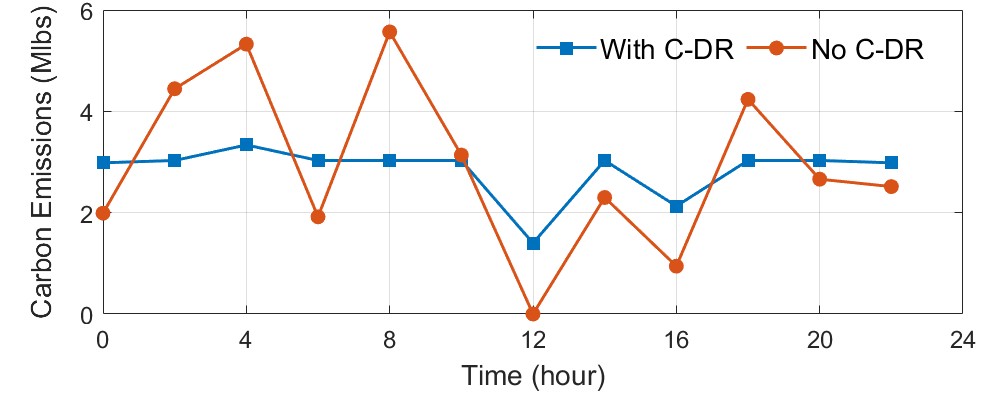}
    \caption{Total system carbon emissions over time under the power dispatch schemes with and without the integration of the C-DR mechanism.}
    \label{fig:emi}
\end{figure}

\begin{figure}
    \centering
    \includegraphics[scale=0.25]{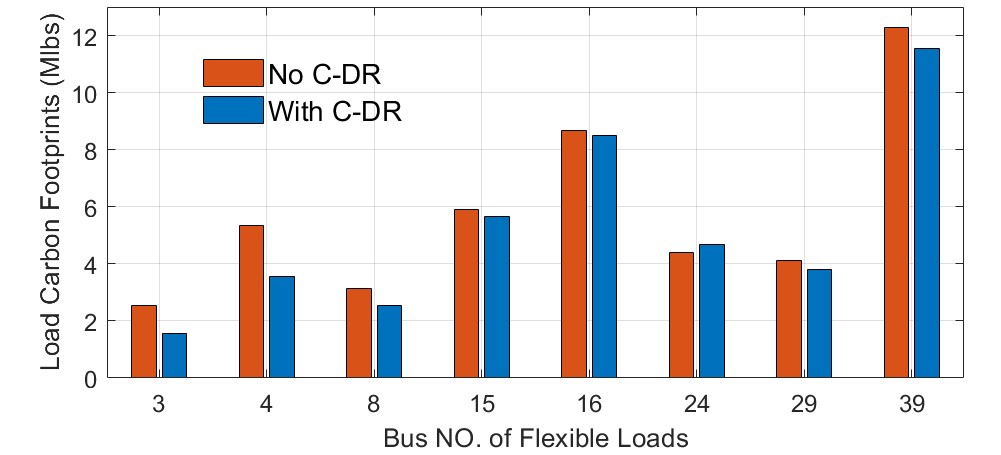}
    \caption{Total carbon footprints of flexible loads with and without implementing the C-DR mechanism.}
    \label{fig:flexible}
\end{figure}

 The total system emissions over time are presented in Figure \ref{fig:emi}. It is observed that the daily system emission curve, when employing the C-DR mechanism, appears flatter compared to when it is not utilized. This phenomenon arises from the behavior of users under C-DR, wherein they adapt their consumption patterns to align with the time-varying nodal carbon intensities. Specifically, users consume more during lower carbon intensity periods, and conversely, reduce consumption during periods of higher carbon intensity. Consequently, this results in diminished temporal fluctuations in system emissions.
In Figure \ref{fig:flexible}, we present a comparison of the total carbon footprints associated with flexible loads. The results reveal that users can generally diminish their carbon footprints by adopting the C-DR mechanism. However, the extent of carbon footprint reduction varies among users. This discrepancy arises from the dynamic adjustment of power dispatch by the grid operator, resulting in different changes in nodal carbon intensities across the power network. Consequently, the decarbonization outcomes differ across users at distinct nodes.

\begin{figure}
    \centering
    \includegraphics[scale=0.25]{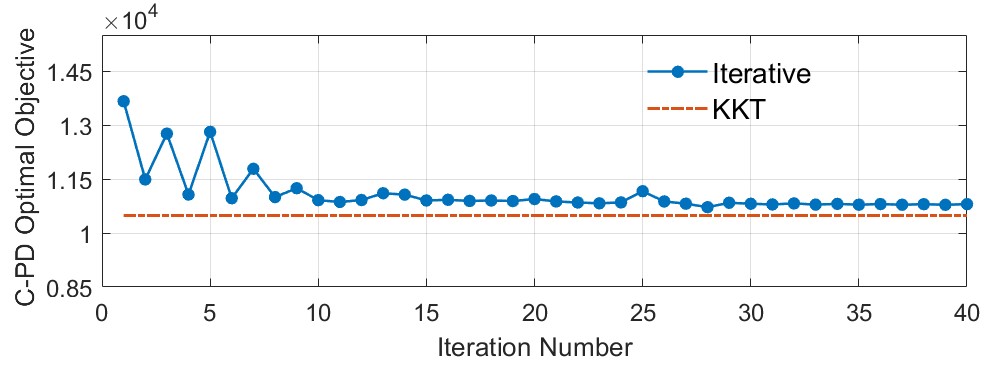}
    \caption{Convergence of the iterative solution algorithm (Algorithm \ref{alg:ite}).}
    \label{fig:algo}
\end{figure}

Figure \ref{fig:algo} illustrates the convergence results of the iterative solution algorithm (Algorithm \ref{alg:ite}) for solving the bi-level C-PD optimization model \eqref{eq:cpd}, compared with the optimal objective value obtained by the KKT reformulation method. For
constraint \eqref{eq:addc} in Algorithm \ref{alg:ite}, we set the parameters $M_l = 5\times 10^3$ and $\alpha_l = 1.5$. As shown in Figure \ref{fig:algo}, oscillations are observed initially and then 
Algorithm \ref{alg:ite} converges rapidly within tens of iterations. It demonstrates the effectiveness of constraint \eqref{eq:addc} in attenuating oscillations and ensuring algorithmic convergence.

\section{Conclusion} \label{sec:conclusion}

This paper studies the carbon-aware demand response (C-DR) paradigm where individual users interact with power grids to curtail their carbon footprints through optimal scheduling of flexible load devices. The time-varying nodal carbon intensities are used as granular grid emission signals to guide users' load adjustment. 
It results in a C-DR mechanism that relates individual users' load profiles to the time-varying nodal carbon intensities.
This C-DR mechanism is then integrated into the optimal power dispatch model for low-carbon power system operation, built upon the carbon-aware optimal power flow (C-OPF) methodology. A centralized KKT reformulation algorithm and an iterative solution algorithm are proposed to solve the resultant bi-level optimal power dispatch model. 
The numerical tests on a modified IEEE 
 39-bus system demonstrate that the C-DR-embedded power dispatch model can effectively utilize the power flexibility on the user side to reduce power system carbon emissions and operational costs. 
Future work encompasses two key aspects: 1) considering the uncertainty in users' behaviors, and 2) refining the solution algorithms for solving the bi-level power dispatch optimization model.






\bibliography{ref}

\end{document}